\documentclass[english,polish,american]{appolb}
\usepackage{caption}
\usepackage{graphicx}
\usepackage{amsmath}
\usepackage{epstopdf}

\usepackage{babel}
\begin{document}

\title{Semi-classical, microscopic approach to the liquid drop model - a
possible way of the description of heavy ion reaction.}

\author{Zbigniew Sosin$^{}$\thanks{\mbox{\emph{E-mail~address}:~ufsosin@cyf-kr.edu.pl (Zbigniew Sosin).}}          ,
  Jinesh Kallunkathariyil\\
\footnotesize 
\textit{M. Smoluchowski Institute of Physics, Jagiellonian University, Reymonta
4, PL-30059 Krakow, Poland}\\
}

\maketitle

\abstract{
An isospin and spin dependent form of the equation of state for nuclear
matter is presented. This form is used for the description of nucleon
interaction in a new dynamic model. Preliminary calculations show
that the new approach makes possible predicting the alpha structures
appearing in the case of the ground state even-even nuclei.
}

\section{Introduction}

Although nuclei are close to 55 orders of magnitude lighter than a
typical neutron star, it seems that the state of the material comprising
of the above two objects can be determined by the same equation, which
is the equation of the state (EOS) of nuclear matter. In this case EOS
is defined as the average energy per baryon expressed as a function
of thermodynamic variables. It seem that the above statement
can be more easily justified for heavy nuclei but, as can be shown,
using an appropriate form of the equation of state can give good results
also for the description of the ground states for lighter nuclei.
We arrived at these conclusions using semi-classical, microscopic version
of the liquid drop model in which dynamics is governed by EOS. 

The description of the nuclear system dynamics, which uses the concept
of the equation of state, can be found in the reaction models referring
to the liquid drop model (LDM) \cite{Blocki78}. In this description,
it is assumed that the energy of the system is determined mainly by
the following three components:

- the volume energy defined by the kinetic energy related to internal,
fermionic motion and respective potential interactions for infinite
nuclear matter in equilibrium,

- the surface energy, which in classical description could be treated
as a result of the surface tension action,

- the Coulomb energy associated with the proton charge.

Such models typically assume one body dissipation and no compression
of the nuclear matter. In such an approach the volume energy together
with energies which are corrections related to a not equal number
of protons and neutrons and a non-zero spin of the system may be treated
as the $0^{th}$ term of the expansion of the energy density function
(given by EOS). No matter compression assumed in this model introduces
additional constraints on the global coordinates which describe the
system shape and its evolution.

In our microscopic description of the system evolution, we take into
account also the next term of the expansion of the equation of state
as a function of the density of nuclear matter. The presented approach
treats the local density of the nuclear matter as a result of the aggregation
of partial densities given by Gaussian packets which represent the
nucleons forming the system. As a result, the local density of a matter
is explicitly given by the position of these packages and their variance.
In our approach we describe the volume energy as the respective functional
(volume integral of the energy density given by the equation of state
multiplied by the density of the nuclear matter).

The nuclear EOS describes equilibrated, infinite nuclear matter and
is used in the calculations of the properties of the astronomical objects
as, for example, neutron stars or supernovae \cite{Latt_01}. In such
large objects, the surface effect in the description of the total
energy is negligible. For small objects, such as atomic nuclei, the
energy due to interactions at the surface, cannot be ignored and in
such case the systen energy (the Hamiltonian values) must have additional 
contributions, corresponding to this additional surface energy. In the presented
approach, the surface energy is related to the dispersion of the energy
density in the nucleon's environment and also to the increase of the
diffuseness of the nucleon momentum caused by nucleon location on
the surface (reduce the uncertainty given by the surface interactions
as described in \cite{Sos_2010}). 

In the interaction description which is applied in the model presented
here one can distinguish ingredients related to the volume, surface
and Coulomb energy, which are the essence of the description used in the LDM
model .

In our approach, all the aforementioned types of energy are functional
expressed by integrals determined by the distribution of nuclear matter.
Since these densities are here uniquely determined by the position
and other parameters of wave packets (spin, isospin and variance),
the description can be interpreted as the Microscopic LDM model (MLDM).

As we know, the equation of state of nuclear matter is still not well
established and its determination is one of the most important tasks
faced by experimental and theoretical nuclear physics. In the presented
approach, the equation of state plays a very important role and in
this way the described model has an opportunity to contribute in EOS
investigation. It is also important to use a convenient form of EOS
suitable for efficient description of the heavy ion reactions as well
as for the description of the ground state properties of the reacting nuclei.
This form of EOS should fulfil the following requirements:

1. The compression part of the equation of state, in the neighbourhood
of saturation density $\rho_{0}$ (where $\rho_{0}$ is the density
of the matter which is equilibrated in isospin and spin and for which
the density of energy reaches a minimum) can be written as the sum
of the two types of energy: 

- $e_{0}$: energy related to symmetric nuclear matter (SNM); 

- $e_{sym}$: energy produced by the nuclear matter symmetry disorders,
which is called a symmetry energy (see section 2). 

It turns out that adopting certain simplifying assumptions
(see appendix A), these energies can be expressed as a function of
the total density $\rho$ . The expectation which is in line with
a mean field theory (e.g. with the Skyrme-Hartree-Fock model) determine
the zero energy density in case where density of matter tends to zero
(it is in contrast with some other models like, for instance, an idealized
alpha-mater picture \cite{John80}).

2. For the saturation density we require: 

- a defined value of the energy density;

- a defined values of the first and the second derivative (slope and
curvature) of the energy density (defined by EOS);

and for density $\rho\rightarrow0$ the additional requirement is
\foreignlanguage{english}{$e(\rho)\rightarrow0$} 

then the simplest form of a polynomial, which can meet all these requirements
can be expressed as:
\begin{equation}
e=\alpha\rho+\beta\rho^{2}+\gamma\rho^{3}\label{eq:wzor}
\end{equation}

The method of determining the coefficients alpha, beta, gamma is given
in Section 2. and Appendix A. As it is shown \cite{Sos_2010}, form
(\ref{eq:wzor}) is suitable for different theoretical descriptions
of the EOS. It is also worth noting that this form of the EOS allows us
to find the respective mean energy values in an analytical way (in
case when the density is expressed by partial density, taken as the
Gaussian functions).

A modified equation of state in a form suitable for MLDM calculations
is defined in the next section. In the third section we will present
the application of the proposed form of the EOS together with an additional
surface energy term in MLDM model. The obtained preliminary results
using this model are related to the properties of nuclei in their ground
states and are presented in Section 4. Since in our description we
use EOS, in which energy symmetry is associated with spin and isospin
asymmetry (see next section), the geometric structure defined by the
position of the centres of the wave packets can be interpreted as
clusters. For the nuclei with even and equal number of neutrons and
protons these clusters appear as alpha particles. In Section 5 we
present such alpha structures, which appear in model calculations
for light nuclei.

In the last chapter conclusions and future plans are presented.

\section{An equation of state given by the third-degree polynomials}

The concept of the EOS of nuclear matter in the case
of two-component proton-neutron gas has been discussed in many articles.
This equation as well as a form based on the use of
of the third-degree polynomials is described in \cite{Sos_2010} and works
cited therein. In the next section we present the ground state properties
of nuclei which are obtained by the microscopic,density-functional
approach. The EOS plays a key role in this approach.

This model is designed to explore the dependence of the dynamics on
wave packets, isospin and spin of nucleons, which these packets
represent. By isospin dependence, we mean the dependence on the charge
of the nucleon (third component of isospin vector). 

If the isospins, spins and variances of wave packets are the variables
affecting the mutual interaction of nucleons, then appropriate equations
of state must describe a gas consisting of four components: 

-protons with spin up - density $\rho_{p\uparrow}$

-protons with spin down - density $\rho_{p\downarrow}$

-neutron with spin up - density $\rho_{n\uparrow}$

-neutron with spin down, density $\rho_{n\downarrow}$

In our considerations, by 'nucleons with spin up' or 'nucleons with
spin down' we mean nucleons having spin projections 'up' or 'down'
on the chosen quantization axis. Now the EOS can be formally written:
\begin{equation}
e=e(\rho_{p\uparrow},\rho_{p\downarrow},\rho_{n\uparrow},\rho_{n\downarrow})\label{eos}
\end{equation}
According to symmetries characterizing the nuclear interaction, the
system energy is conserved during mutual exchange of neutrons and
protons, and also when changing projections of the spin of all
particles to the opposite projections. To allow us to use these symmetries
we define new coordinates:

\begin{equation}
\xi=\frac{\rho-\rho_{0}}{\rho_{0}}\label{eq:ksi}
\end{equation}

\begin{equation}
\delta=\frac{\rho_{n}-\rho_{p}}{\rho}\label{eq:del}
\end{equation}
\begin{equation}
\eta_{n}=\frac{\rho_{n\uparrow}-\rho_{n\downarrow}}{\rho}\label{eq:n_n}
\end{equation}
\begin{equation}
\eta{}_{p}=\frac{\rho_{p\uparrow}-\rho_{p\downarrow}}{\rho}\label{eq:n_p}
\end{equation}
 where $\rho$ is the total nuclear matter density, and $\rho_{0}$
is the density of the isospin and spin equel to zero matter at saturation.
We find that during the above-mentioned operation (mutual conversion
of neutrons into protons and the change all the nucleon spin projections)
the sign of the coordinates: neutron-proton asymmetry $\delta$ and
spin asymmetry $\eta_{n}$ and $\eta_{p}$ are respectively changed.
Therefore in order to assure energy conservation, in the expression
which defines the system energy values of $\delta$ , $\eta_{n}$
and $\eta_{p}$ can occur only as products with the respective even
number of factors. Using these symmetries one can show (see Appendix
A) that the equation of state can be written with the components having
an extended form of symmetry energy:
\[
e=e_{00}+\frac{K_{0}}{18}\xi^{2}+
\]
\[
\delta^{2}\left(e_{I0}+\frac{L_{I}}{3}\xi+\frac{K_{I}}{18}\xi^{2}\right)+
\]

\[
\left(\eta_{n}^{2}+\eta_{p}^{2}\right)\left(e_{ii0}+\frac{L_{ii}}{3}\xi+\frac{K_{ii}}{18}\xi^{2}\right)+
\]
\begin{equation}
2\eta_{n}\eta_{p}\left(e_{ij0}+\frac{L_{ij}}{3}\xi+\frac{K_{ij}}{18}\xi^{2}\right)\label{eq:eos_4}
\end{equation}

The first two components of the sum (\ref{eq:eos_4}) represent a
well-known form of the equation of state. The first component describes
the matter in a balanced system (zero isospin and spin), where $K_{0}$
is the coefficient of compressibility of nuclear matter and the second
one describes the isospin interaction is given in the following form:
\begin{equation}
e_{I}=e_{I0}+\frac{L_{I}}{3}\xi+\frac{K_{I}}{18}\xi^{2}\label{eq:esym_I}
\end{equation}
which is called the symmetry energy for which $e_{I0}$ is the Wigner
constant and the coefficients $L_{I}$ and $K_{I}$ are the slope
and curvature respectively.

In equation (\ref{eq:eos_4}), by analogy to the isospin symmetry
energy, one can distinguish the spin symmetry energies for neutrons
and protons separately: 
\begin{equation}
e_{ii}=e_{ii0}+\frac{L_{ii}}{3}\xi+\frac{K_{ii}}{18}\xi^{2}\label{eq:esym_ii}
\end{equation}
and energy:
\begin{equation}
e_{ij}=e_{ij0}+\frac{L_{ij}}{3}\xi+\frac{K_{ij}}{18}\xi^{2}\label{eq:esym ij}
\end{equation}
 for the mutual, spin interaction of protons and neutrons. Factors
develop spin symmetry energy: $e_{ii0}$, $L_{ii}$ and $K_{ii}$,
(constant slope and curvature) with indices $ii$, describe the energy
related to the proton or neutron gas. While the respective indices
$ij$ indicate that the symmetry energy refers to the mutual interaction
of protons and neutrons.

All types of the symmetry energy describe the influence of the spin and
isospin polarization of the matter on the average energy of the nucleon
(EOS). According to the observations presented in the introduction
of work, we assume that the energy associated with the SNM as well
as all types of symmetry energy should disappear as the density of
matter tends to zero. Therefore, to ensure zero values for all energies
occurring in equation (\ref{eq:eos_4}) at $\rho=0$ and to get the
proper slope and curvature of the saturation density $\rho_{0}$ we
use the form of the third-degree polynomial proposed in \cite{Sos_2010}.
As mentioned in the \cite{Sos_2010}, if the density of matter is described
by the sum of Gaussian distributions, then such a form of the EOS
allows us to calculate the average energy analytically. 

Any type of energy found in (\ref{eq:eos_4}) may be written as:
\begin{equation}
e_{k}=\alpha_{k}\rho+\beta_{k}\rho^{2}+\gamma_{k}\rho^{3}\label{eq:form}
\end{equation}
and the index $'k'$ can take values $0$, $I$, $ii$, and $ij$.
Here the coefficients $\alpha_{k},\:\beta_{k},\:\gamma{}_{k}$ are
determined by the relations:

\begin{equation}
\alpha_{k}=\left(3e_{k0}-\frac{2}{3}L_{k}+\frac{K_{k}}{18}\right)/\rho_{0}\label{eq:al_k}
\end{equation}

\begin{equation}
\beta_{k}=-\left(3e_{k0}-L_{k}+\frac{K_{k}}{9}\right)/\rho_{0}^{2}\label{eq:be_k}
\end{equation}

\begin{equation}
\gamma_{k}=\left(e_{k0}-\frac{L_{k}}{3}+\frac{K_{k}}{18}\right)/\rho_{0}^{3}\label{eq:ga_k}
\end{equation}

\section{Application of the proposed EOS form to the MLDM calculations}

The proposed MLDM, similarly like in models \cite{Aich91},\cite{Ono_92},\cite{Feld_90},
describe the time evolution of the wave function represented by the
product of $M$ Gaussian wave packets, which represent the nucleons
forming the system :
\begin{equation}
\Phi={\displaystyle \prod_{k=1}^{M}}\,^{k}\phi_{I_{k}S_{k}}\label{eq:wf}
\end{equation}
\begin{equation}
\,^{k}\phi_{I_{k}S_{k}}=\frac{1}{\left(2\pi\sigma_{k}^{2}(r)\right)^{3/4}}\exp\left(\frac{-\left(\mathbf{r}_{k}-\left\langle \mathbf{r}_{k}\right\rangle \right)^{2}}{4\sigma_{k}^{2}(r)}+\frac{i}{\hbar}\mathbf{r_{k}}\left\langle \mathbf{p}_{k}\right\rangle \right)\label{eq:wf_k}
\end{equation}
where $\sigma_{k}^{2}(r)$, $\left\langle \mathbf{r}_{k}\right\rangle $,
$\left\langle \mathbf{p}_{k}\right\rangle $, are the width (the position
variance of the $k$-th nucleon) of a Gaussian wave functions, and
centres of its position and associated momentum for each of the $M$ nucleons.
Also every partial wave function (\ref{eq:wf_k}) has a label $I_{k}=n\;\mathrm{or}\; I_{k}=p$
and $S_{k}=\uparrow\;\mathrm{or}\; S_{k}=\downarrow$, informing us
about isospin and spin associated with the given nucleon and determine
the kind of density $\left(\rho_{p\uparrow},\rho_{p\downarrow},\rho_{n\uparrow},\rho_{n\downarrow}\right)$
to which given wave packed participate. In our approach the variables
$\left\langle \mathbf{r}_{k}\right\rangle $, $\left\langle \mathbf{p}_{k}\right\rangle $
and $\sigma_{k}^{2}(r)$ are time-dependent parameters describing
the wave functions. In addition, we assume that with every wave packet
for each of $M$ nucleons, isospin and spin are related and that
they remain fixed during the interaction (in this way index $k$
inform us about the parameters of the given nucleon). 

Equations of motion of variables $\left\langle \mathbf{r}_{k}\right\rangle $,
$\left\langle \mathbf{p}_{k}\right\rangle $ and $\sigma_{k}^{2}(r)$
are derived using the time-dependent variational principle (see e.g.
\cite{Ritz1909}) based on the action minimization. For this purpose,
we define the action as:
\begin{equation}
S=\intop_{t_{1}}^{t_{2}}L\left(\Phi,\Phi^{*}\right)dt\label{eq:action}
\end{equation}
with the Lagrange functional given as:
\begin{equation}
L=\left\langle \Phi\left|i\hbar\frac{d}{dt}-H\right|\Phi\right\rangle =\left\langle \Phi\left|i\hbar\frac{d}{dt}\right|\Phi\right\rangle -\left\langle \Phi\left|H\right|\Phi\right\rangle \label{eq:Lagrange}
\end{equation}

Now we will concentrate on the determination of the average value of
Hamiltonian$\left\langle \Phi\left|H\right|\Phi\right\rangle $. As
usual the Hamiltonian can be written as a sum of the kinetic energy
$T$ and potential interaction $V$. To determine the average kinetic
energy we assume that each wave packet has a momentum dispersion $\sigma_{k}(p)$
around its mean value $\left\langle \mathbf{p}_{k}\right\rangle $,
($\sigma_{k}(r)\sigma_{k}(p)=\hbar/2$ is assumed). Then the average
kinetic energy associated with the system is given by:

\begin{equation}
\left\langle \Phi\left|T\right|\Phi\right\rangle ={\displaystyle \sum_{k=1}^{k=M}\left[\frac{\left\langle \mathbf{p}_{k}\right\rangle ^{2}}{2m}+\frac{3\sigma_{k}^{2}(p)}{2m}\right]}\label{eq:Ekin}
\end{equation}
We assume also that the interaction between nucleons can be described
by the potential given in the form:
\begin{equation}
V\left(\mathbf{\left\{ r_{k}\right\} },\left\{ \sigma_{k}(r)\right\} \right)=V_{N}\left(\mathbf{\mathbf{\left\{ r_{k}\right\} }},\left\{ \sigma_{k}(r)\right\} \right)+V_{S}\left(\mathbf{\mathbf{\left\{ r_{k}\right\} }},\left\{ \sigma_{k}(r)\right\} \right)+V_{Coul}\left(\mathbf{\mathbf{\left\{ r_{k}\right\} }},\left\{ \sigma_{k}(r)\right\} \right)\label{eq:Pot_tot}
\end{equation}
where $V_{N}$ describes the nuclear interaction, $V_{S}$ describes
the modification of the interaction induced by changing the density
of matter around the nucleon (surface energy) and $V_{Coul}$ consists
of the Coulomb interaction. If the nuclear matter is characterized
by non zero spin polarization then, in the space a magnetic field
exist. This field can also be generated by the motion of protons.
In such a case the Hamiltonian should be corrected for the magnetic
interaction. 

Now the average value of the Hamiltonian can be expressed by:
\begin{equation}
\left\langle \Phi\left|H\right|\Phi\right\rangle ={\displaystyle \sum_{k=1}^{k=M}\left[\frac{\left\langle \mathbf{p}_{k}\right\rangle ^{2}}{2m}+\frac{3\sigma_{k}^{2}(p)}{2m}\right]+\left\langle \Phi\left|V_{N}\right|\Phi\right\rangle +\left\langle \Phi\left|V_{S}\right|\Phi\right\rangle +\left\langle \Phi\left|V_{Coul}\right|\Phi\right\rangle }+\left\langle \Phi\left|M\right|\Phi\right\rangle \label{eq:Hamiltonian}
\end{equation}
 In our further considerations we ignore magnetic interaction. The
crucial assumption of the present approach is that the energy of the fermionic
internal motion can be described by $\frac{3\sigma_{p_{k}}^{2}}{2m}$,
so the sum 
\begin{equation}
B_{V}=\sum_{k=1}^{k=M}\frac{3\sigma_{k}^{2}(p)}{2m}+\left\langle \Phi\left|V_{N}\right|\Phi\right\rangle =\intop e(\rho,\delta,\eta_{n},\eta_{p})\rho(\mathbf{r})\, d^{3}\mathbf{r}\label{eq:B_V}
\end{equation}
can be treated as a volume term of binding energy. As we know, the
ground state energy of the nuclear matter is described by the EOS, which
considers the energy density as a sum of the kinetic energy, associated
with the internal fermionic motion, and energy given by the potential
interactions. Therefore in our approach we do not have to
find the appropriate distribution of the nucleon momenta, provided
that a correct description is given by the selection of appropriate
EOS parameters. Such an EOS parametrization replaces the potential
parametrization normally used in such approaches.

As in \cite{Sos_2010}, the surface energy is defined as 
\begin{equation}
B_{surf}=\left\langle \Phi\left|V_{S}\right|\Phi\right\rangle =s_{0}\sum_{i=1}^{i=M}\frac{\sigma_{i}(e)}{\sigma_{i}^{2}(r)}\label{eq:surface_energy}
\end{equation}
where in this semi-empirical part $s_{0}$ is the surface energy coefficient.
In this formula, the variance $\sigma_{k}(e)$ denotes the variance
of energy $e(\rho,\delta,\eta_{n},\eta_{p})$ with respect to the
probability $^{k}\!\rho(\mathbf{r})$ of finding the $k$-th nucleon:
\begin{equation}
\sigma_{k}(e)=\left[\intop\left(\bar{e}_{k}-e(\rho,\delta,\eta_{n},\eta_{p})\right)^{2}\cdot^{k}\!\!\!\rho(\mathbf{r})d^{3}\mathbf{r}\right]^{\frac{1}{2}}\label{eq:sig_e}
\end{equation}
where $\bar{e}_{k}$ is the average energy given by: 
\begin{equation}
\!\bar{e}_{k}=\intop e(\rho,\delta,\eta_{n},\eta_{p})\cdot^{k}\!\!\!\rho(\mathbf{r})d^{3}\mathbf{r}\label{eq:ave_e}
\end{equation}
In appendix A we describe the method of calculation used for variance
$\sigma_{k}(e)$ calculations.

Let us discuss briefly the description of the surface energy given
by formula (\ref{eq:surface_energy}). In our approach we consider
drops of nuclear matter shown in Figure 1.
\begin{figure}[h]
\includegraphics[width=12cm]{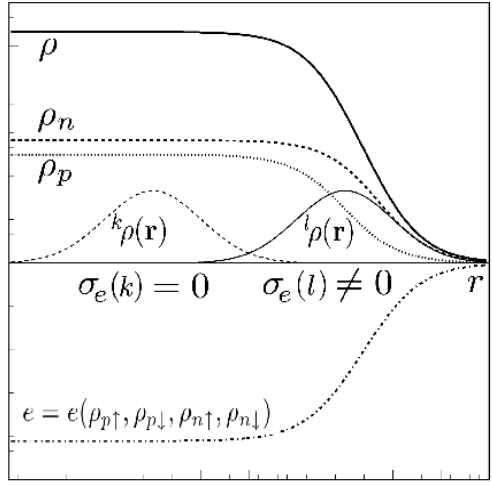}

\caption{An example of the position of the partial density function and associated
energy variance. As one can see for a position close to the center,
variance of the energy (\ref{eq:sig_e}) vanishes. See text for more
details. }
\end{figure}
The density of matter $\rho(r)=\rho_{n}(r)+\rho_{p}(r)$ is a function
of the distance $r$ and $\rho_{n}(r)=\rho_{n\uparrow}+\rho_{n\downarrow}$
and $\rho_{p}(r)=\rho_{p\uparrow}+\rho_{p\downarrow}$. If the $k^{th}$
packet is in the region of  fixed energy $e(\rho,\delta,\eta_{n},\eta_{p})$,
then its variance $\sigma_{k}(e)$, associated with this package $^{k}\!\rho(\mathbf{r})$,
is equal to zero. If the package $^{l}\!\rho(\mathbf{r})$ is in the
change of the energy, then the variance $\sigma_{l}(e)$ is greater
than zero. According to the formula (\ref{eq:surface_energy}), the surface
energy associated with the $i^{th}$ packet is inversely proportional
to the uncertainty of location, which means that it is proportional
to the indetermination of the package momentum. Now the formula (\ref{eq:surface_energy})
can be understood as follows. Variation of energy $\sigma_{i}(e)$
(and associated forces) describes the possibility of localization
of the $i^{th}$-th nucleon in case of its entering the alternating
field. Such localization forces result in an increase of indetermination
of the momentum and an increase of the average kinetic energy associated
with the $i^{th}$ package .

The last term in (\ref{eq:Pot_tot}) is the Coulomb energy (for protons
only) and is given in the form: 
\begin{equation}
\left\langle \Phi\left|V_{Coul}\right|\Phi\right\rangle =\frac{1}{2}{\displaystyle \sum_{k\neq l}}\intop{}^{k}\!\rho(\mathbf{r})\frac{e^{2}}{\left|\mathbf{r-r'}\right|}\:{}^{l}\!\rho(\mathbf{r'})d^{3}\mathbf{r}d^{3}\mathbf{r}'\label{eq:Coul_energy}
\end{equation}
This integral can be calculated analytically as (see eg. \cite{Papa_01}):

\begin{equation}
e^{2}{\displaystyle \sum_{k\neq l}}\frac{\mathrm{erf}\left(\frac{r_{kl}}{\sqrt{2}\sigma_{kl}(r)}\right)}{r_{kl}}\label{eq:pot_erf}
\end{equation}
where $r_{kl}=\left|\left\langle \mathbf{r}_{k}\right\rangle -\left\langle \mathbf{r}_{l}\right\rangle \right|$
and sigma $\sigma_{kl}(r)=\sqrt{\sigma_{k}^{2}(r)+\sigma_{l}^{2}(r)}$.

Finally, the expected form of the Hamiltonian can be written in a
form which is analogous to the description given by the liquid drop
model (LDM) :
\begin{equation}
\left\langle \Phi\left|H\right|\Phi\right\rangle =\sum_{k=1}^{k=M}\frac{\left\langle \mathbf{p}_{k}\right\rangle ^{2}}{2m}+B_{V}\left(\left\{ \left\langle \mathbf{r}_{k}\right\rangle \right\} ,\left\{ \sigma_{k}(r)\right\} \right)+B_{S}\left(\left\{ \left\langle \mathbf{r}_{k}\right\rangle \right\} ,\left\{ \sigma_{k}(r)\right\} \right)+V_{Coul}\left(\left\{ \left\langle \mathbf{r}_{k}\right\rangle \right\} ,\left\{ \sigma_{k}(r)\right\} \right)\label{eq:ave_Ham}
\end{equation}

Based on the Euler-Lagrange equations one can find the corresponding
equations of motion for coordinates $\left\langle \mathbf{p}_{k}\right\rangle ,\left\langle \mathbf{r}_{k}\right\rangle $
and $\sigma_{k}(r)$ (see Euler-Lagrange equations). This will
be described in a forthcoming paper.

Here, as the first test of the new form of the equation of state,
based on the MLDM model, we will try to describe the ground state
properties of a selected set of nuclei. In our very preliminary approach
we will focus on binding energies and nuclear charge radii (RMS).

\section{The MLDM calculations for describing the ground state properties
of nuclei }

In the MLDM model the ground state of a nucleus is a many-body state
which is an absolute minimum with respect to variations of $\left\langle \mathbf{r}_{k}\right\rangle ,\:\sigma_{k}(r)$.
Additionally, in the ground state, for every $k$ the average momentum
$\left\langle p_{k}\right\rangle =0$ , which means that for MLDM
the particle will be essentially at rest and the system corresponds
to a solid one. In order to determine the nuclear ground state wave function
the following procedure of searching the set $\left\{ \left\langle \mathbf{r}_{k}\right\rangle ,\sigma_{k}(r)\right\} $
is used:

1. In a limited space we choose a random position of the wave packet
together with a certain initial variance. In order to accelerate
calculations for even-even nuclei we can also assume a correlation
(mating) for packets describing nucleons of the same type and
differing only in the direction of the spin projection. In such a
case we assume that the positions of centres and the variances of
a spin pair are equal.

2. Using a selected set of parameters describing the interaction (type
EOS) we search for such values of these parameters for which the
energy of the system is minimal. Here, a variety of methods may be
used. One should also be careful to avoid emerging local minima. Wave
functions found in this way are characterized by certain symmetries,
which may be helpful in assessing the resulting minimum.  In the next
chapter we will discuss types of symmetry, emerging especially for
even-even nuclei with an equal numbers of protons and neutrons.

To use the presented model for calculation of ground state properties
of nuclei, we have to select specific values of parameters describing
the equation of state. For this purpose a global search for optimal
parameters values should be used. This will be done in the next paper,
here we apply only a very simplified search method based on following
assumptions:

In the first step we try to determine the $\rho_{0},\: e_{00},\: K_{0}$
and $s_{0}$ parameters . In our preliminary estimations, we assume
that $\rho_{0}=0.159$ $\textrm{nucl/f\ensuremath{m^{3}}}$ and $K=300$
MeV, which have to be in an agreement with the experiment and theoretical calculations.
Then we use the binding energies and RMS radii for $^{4}He$ and $^{12}C$
nuclei to determine $e_{00}$ and $s_{0}$ by the trial and error
method. In this way we obtain $e_{00}=-12.9$ MeV and $s_{0}=0.09\:\textrm{f\ensuremath{m^{2}}}$.
For these nuclei, due to the matter gathering in clusters which are
equilibrated in spin and isospin coordinates, corrections arising
due to the symmetry energies appear to be negligibly small. Note that
the obtained parameters $e_{00}$ and $s_{0}$, differ from standard
values (in particular, the typical value of $e_{00}=-16$ MeV). This
is the result of their assessment for  very light nuclei. In the future
it will be necessary to determine these parameters based on a larger
database (the set of the energy and radii of the nuclei in their ground
states).

In the next step, we use the binding energy and RMS radii of $^{2}H$,
$^{3}H$, $^{3}He$, to obtain an estimate of the isospin and spin
polarization parameters. Our rough estimate gives $e_{I0}=30$ MeV,
$L_{I}=123$ MeV, $K_{I}=500$ MeV for isospin parameters and $e_{ii0}=65$
MeV, $L_{ii}=300$ MeV, $K_{ii}=300$ MeV for the spin interaction
in gas of neutrons (these same parameters are for  proton gas). 
The mutual spin interaction of protons
and neutrons are described by the parameters $e_{ij0}=-1$ MeV, $L_{ij}=20$
MeV, $K_{ij}=300$ MeV.

With such  aparameters, the dependence of the energies $e_{0}$, $e_{I}$,
$e_{ii}$ and $e_{ij}$, on the density are presented in Figure 2.
\begin{figure}[h]
\includegraphics[width=12cm]{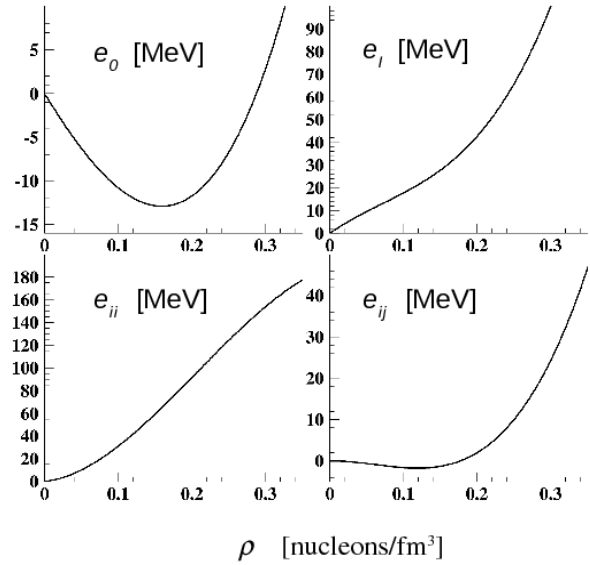}

\caption{\label{fig1-1} The dependence on the density of nuclear matter: $e_{0}$
in an isospin and spin balanced state, the symmetry energy $e_{I}$
for isospin of the system, the symmetry energy $e_{ii}$ for the spin
polarization for proton or neutron matter and the symmetry energy
energy $e_{ij}$ for spin polarization for proton-neutron matter (see
text)}
\end{figure}

In this simple analysis the $^{2}H$, $^{3}H$, $^{3}He$, $^{4}He$
and $^{12}C$ nuclei can be treated as generators of the EOS parameters.
Reproduced by MLDM model, the binding energy and RMS radii are for
this generators arranged together with the experimental data in Table 1

\begin{table}
\caption{ MLDM calculation results (for preliminary EOS parameters selection, preliminary).
Binding energies and RMS charge radii of nuclei are compared with
the experimental data.} 
\small
\begin{tabular}{|c|c|c|c|c|}
\hline 
nuclei & binding  & binding & RMS    & RMS \\
        &  energy  & energy  & charge & charge \\
        &  data    &   MLDM  & radii data & radii MLDM\\
        &  (MeV)   &   (MeV) & (fm)       & (fm)
\tabularnewline
\hline 

$^{2}H$
 &
-1.112
 &
-1.104
 &
2.14 
 &
2.196
\tabularnewline
\hline 

$^{3}H$
 &
-2.827
 &
-2.809
 &
1.759 
 &
1.824 
\tabularnewline
\hline 

$^{3}H$e
 &
-2.572
 &
-2.571
 &
1.945 
 &
1.974 
\tabularnewline
\hline 

$^{4}H$e
 &
-7.074
 &
-7.062
 &
1.676 
 &
1.727 
\tabularnewline
\hline 

$^{12}C$
 &
-7.68 
 &
-7.719
 &
2.47 
 &
2.466
\tabularnewline
\hline
\end{tabular}
\normalsize
\end{table}

For such choice of the EOS parameters (and the coefficient of surface
tension $s_{0}=0.09\:\textrm{f\ensuremath{m^{2}}}$) we obtain predictions
for the binding energies per nucleon and RMS charge radii these are
presented together with experimental data in Table 2.

\begin{table}[h]
\caption{ MLDM calculation results (prediction, preliminary). Binding
energies and RMS charge radii of nuclei are compared with the experimental
data.} 

\small
\begin{tabular}{|c|c|c|c|c|}

\hline 

nuclei & binding  & binding & RMS    & RMS \\
        &  energy  & energy  & charge & charge \\
        &  data    &   MLDM  & radii data & radii MLDM\\
        &  (MeV)   &   (MeV) & (fm)       & (fm)
\tabularnewline
\hline 

$^{16}O$
 & -7.976 &
-7.84 
 &
2.701 
 &
2.613
\tabularnewline
\hline 

$^{20}Ne$
 &
-8.032
 &
-7.908
 &
3.005 
 &
2.774 
\tabularnewline
\hline 

$^{24}Mg$
 &
-8.261
 &
-7.953
 &
3.056 
 &
2.874 
\tabularnewline
\hline 
$^{32}S$
 & %
-8.493
 & %
-7.929
 & %
3.261 
 & %
3.127 
\tabularnewline
\hline 
$^{36}Ar$
 & %
-8.520
 & %
-7.896
 & %
3.39 
 & %
3.227
\tabularnewline
\hline 
$^{20}Ca$
 & %
-8.551
 & %
-7.85 
 & %
3.476 
 & %
3.336 
\tabularnewline
\hline
\end{tabular}
\end{table}
\normalsize

\selectlanguage{american}%
The data in Table 1 and Table 2 should be treated as very preliminary
predictions of the MLDM code based on a roughly chosen equation of
state. For heavier nuclei one can see quite large deviations from
the experimental data. In order to achieve greater consistency with the
description of the experimental data, all EOS parameters should be re-adjusted.

\section{The alpha structure of even-even nuclei with equal number of protons
and neutrons}

The concept of nuclear clusters appeared along with the quantum description
of nuclei (and even before the discovery of the neutron in 1932 by
James Chadwick). In article \cite{Oertzen2006} many experimental
arguments are presented for the possible emergence of alpha clusters,
as a matter substructures in forming nuclei. We present here
only two of them, both taken from \cite{Hafs1938}.

i) The binding energy of even-even nuclei with an equal number of
protons and neutrons appears to be a linear function of the number
of bonds in alpha-particle model, where the number of bonds is equal
to: $k=1$ for $\mathrm{^{8}Be}$, $k=3$ for $\mathrm{^{12}C}$,
$k=6$ for $\mathrm{^{16}O}$,  $k=9$ for $\mathrm{^{20}\mathrm{Ne}}$,
etc. (see Fig. 9). In the cited work this phenomenon is explained
on the basis of the proposed form of the potential interaction between
the alpha particles, however there is no justification for increase
in the number of bounds, with an increasing number of alpha particles.

ii) The binding energy of nuclei with an even number of protons $Z$
and an odd, higher by 1 number of neutrons $N=Z+1$. In this case
we assume that the additional neutron is interacting with multi-centrist
potential defined by the system of the appearing alpha particles.
Let the binding energy for each nuclide $\mathrm{X}$ be defined by
the function $b(\mathrm{X)}$. If the neutron binding energy in $\mathrm{^{5}He}$
is denoted by $B$, then we get the following scheme:

\[
b\left(\mathsf{\mathrm{^{5}He}}\right)-\left[b\left(\mathrm{^{4}He}\right)+b\left(\mathrm{n}\right)\right]=B
\]

\[
b\left(\mathrm{^{9}Be}\right)-\left[b\left(\mathrm{^{8}Be}\right)+b\left(\mathrm{n}\right)\right]=B+\left(R+Q\right)
\]

\[
b\left(\mathrm{^{13}C}\right)-\left[b\left(\mathrm{^{12}C}\right)+b\left(\mathrm{n}\right)\right]=B+2\left(R+Q\right)
\]

\begin{equation}
b\left(\mathrm{^{17}O}\right)-\left[b\left(\mathrm{^{16}O}\right)+b\left(\mathrm{n}\right)\right]=B+3\left(R+Q\right)\label{B_sys}
\end{equation}
where constants $R$ and $Q$ are the result of reasoning, which
below is presented in the case of $^{9}\mathrm{Be}$. If a $\mid\psi_{1}>$
and $\mid\psi_{2}>$ denote the neutron wave functions corresponding
to the interaction with a given alpha particle, then the two-centre
wave function for $\mathrm{^{8}Be}$ can be expressed approximately
as a linear combination of the single center ones ($\mid\psi_{1}>$
and $\mid\psi_{2}>$). Now, the average binding energy of a neutron,
in two-center potential, can be expressed as: 

\begin{equation}
b\left(\mathrm{^{9}Be}\right)-\left[b\left(\mathrm{^{8}Be}\right)+b\left(\mathrm{n}\right)\right]=\frac{<\psi_{1}+\psi_{2}\mid T+V_{1}+V_{2}\mid\psi_{1}+\psi_{2}>}{<\psi_{1}+\psi_{2}\mid\psi_{1}+\psi_{2}>}\label{B_sys_9Be}
\end{equation}

where in the Hamiltonians $H=T+V_{1}+V_{2}$ operator $T$ describes
the kinetic energy and the $V_{1}$ and $V_{2}$ are related to the neutron
interaction with the corresponding alpha particle. As can be seen,
the following expressions occur twice in the numerator:

$<\psi_{1}\mid T+V_{1}\mid\psi_{1}>=B$, which is the neutron binding
energy in $\mathrm{^{5}He}$, and

$<\psi_{1}\mid V_{2}\mid\psi_{1}>=R$, which describes the additional
energy associated with the presence of the second alpha particle,
and

\selectlanguage{english}%
$<\psi_{1}\mid H\mid\psi_{2}>=Q$\foreignlanguage{american}{ as the
energy associated with the exchange process.}

\selectlanguage{american}%
If the wave functions $\mid\psi_{1}>$ and $\mid\psi_{2}>$ are normalized,
then the denominator in (\ref{B_sys_9Be}) is equal to $2(1+<\psi_{1}\mid\psi_{2}>)$.
Usually, it can be assumed that $<\psi_{1}\mid\psi_{2}>=S$ is small
compared to 1, and therefore (\ref{B_sys_9Be}) tends
to the corresponding part of the expression (\ref{B_sys}).

These examples show that the nuclear matter in nuclei has a structure
in which the alpha particles seem to play an important role. 

The report \cite{Oertzen2006} describes also some of alpha cluster
models. Our approach is based on single nucleons and in this way it
is similar to QMD (Quantum Molecular Dynamics) \cite{Aich91}, AMD
(Antisymmetric Molecular Dynamics) \cite{Ono_92} and FMD (Fermionic
Molecular Dynamics) \cite{Feld_90}, in which the dynamics
is based on variational principle.

Since our model includes isospin and spin interactions occurring between
different types of nucleons with densities $\rho_{p\uparrow},\rho_{p\downarrow},\rho_{n\uparrow},\rho_{n\downarrow}$,
nucleons are grouped in order to minimize this additional symmetry energy.
As we know, in the case of fermions this causes the connection of
particles into pairs. For nuclear matter an additional positive energy
is associated with the lack of isospin balance. To minimize both of
these symmetry energies, nucleons create alpha clusters. This phenomenon is
included in our model, so one can observe the alpha structure in the
resulting nuclear matter distribution, particularly for even-even nuclei
with equal proton and neutron numbers. To show this we present in
Figs 3-6 results of calculations for nuclei $^{16}O$, $^{24}Mg$,
$^{36}Ar$ and $^{40}Ca$.
\begin{figure}[h]
\includegraphics[width=12cm]{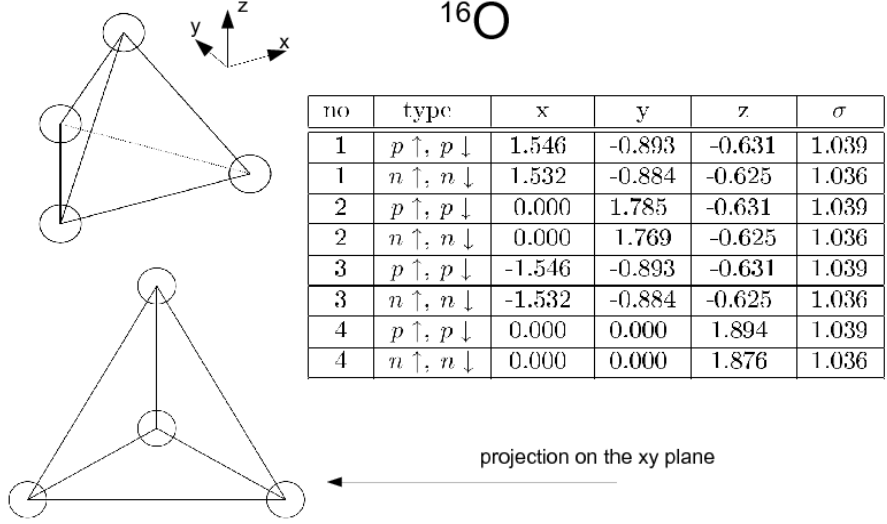}

\caption{The position and the variance of the wave packet for the nucleons
in the nucleus $^{16}O$. In the first column of the table contained
in the figure there are the numbers of groups of nucleons, which are
correlated in the alpha particle. The second column indicates the
spins and type of nucleons creating given pairs of nucleons.The common
position and variance of the two wave packets is presented.}
\end{figure}
\begin{figure}[h]
\includegraphics[width=12cm]{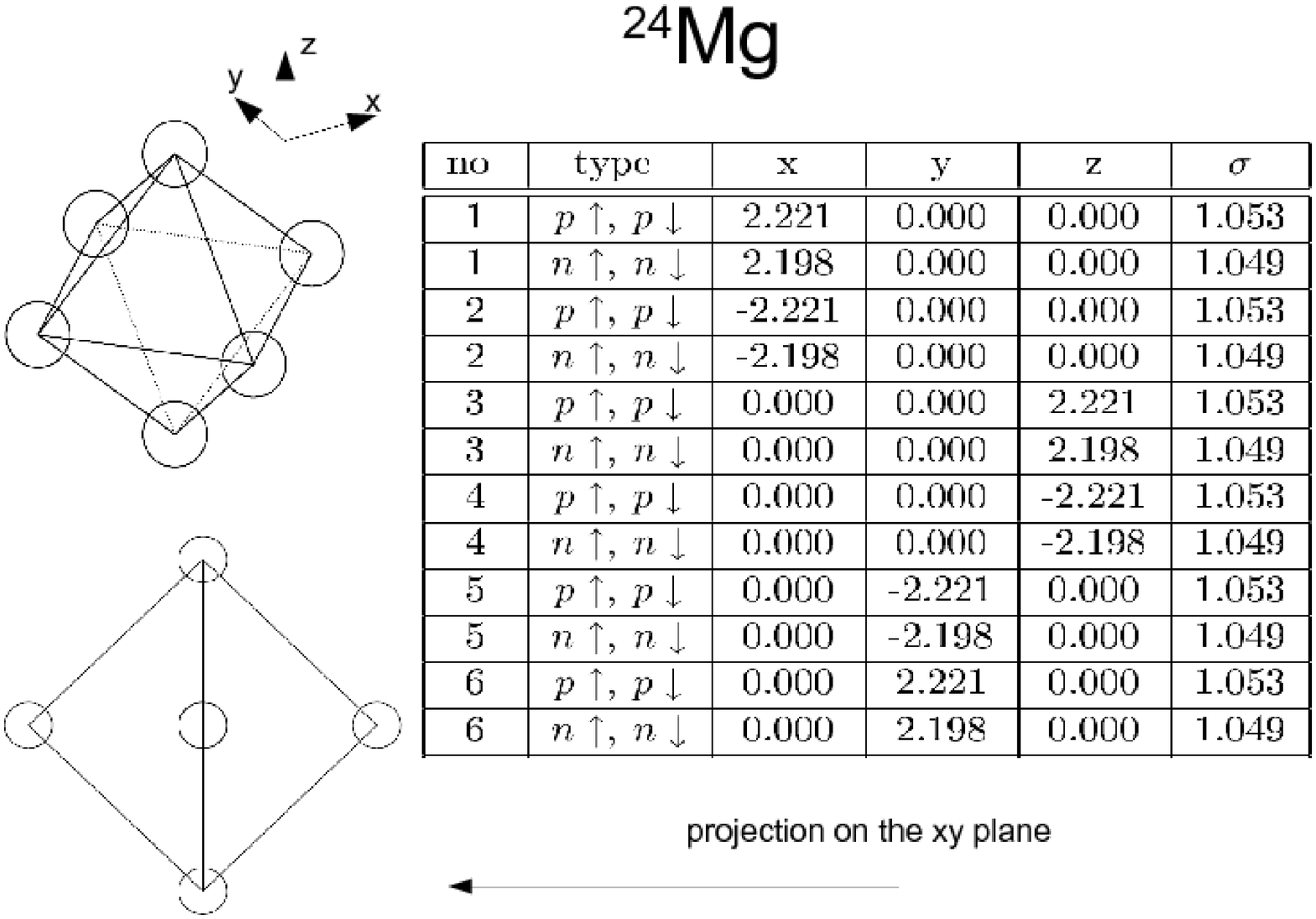}

\caption{The position and the variance of the wave packet for the nucleons
in the nucleus $^{24}Mg$. Notation is the same as in Figure 3.}
\end{figure}

\begin{figure}[h]
\includegraphics[width=12cm]{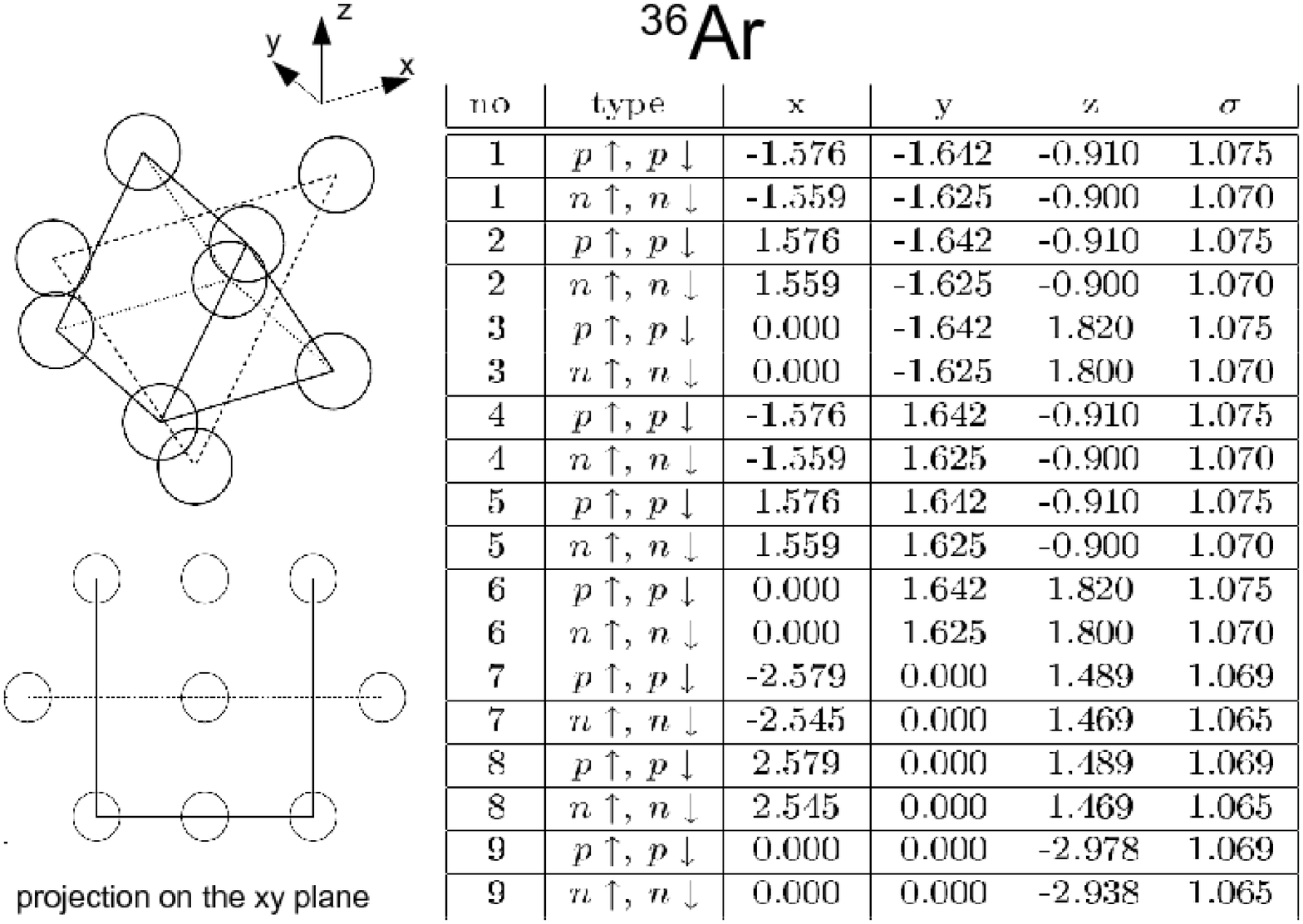}

\caption{The position and the variance of the wave packet for the nucleons
in the nucleus $^{36}Ar$. Notation is the same as in Figure 3.}
\end{figure}
\begin{figure}[h]
\includegraphics[width=12cm]{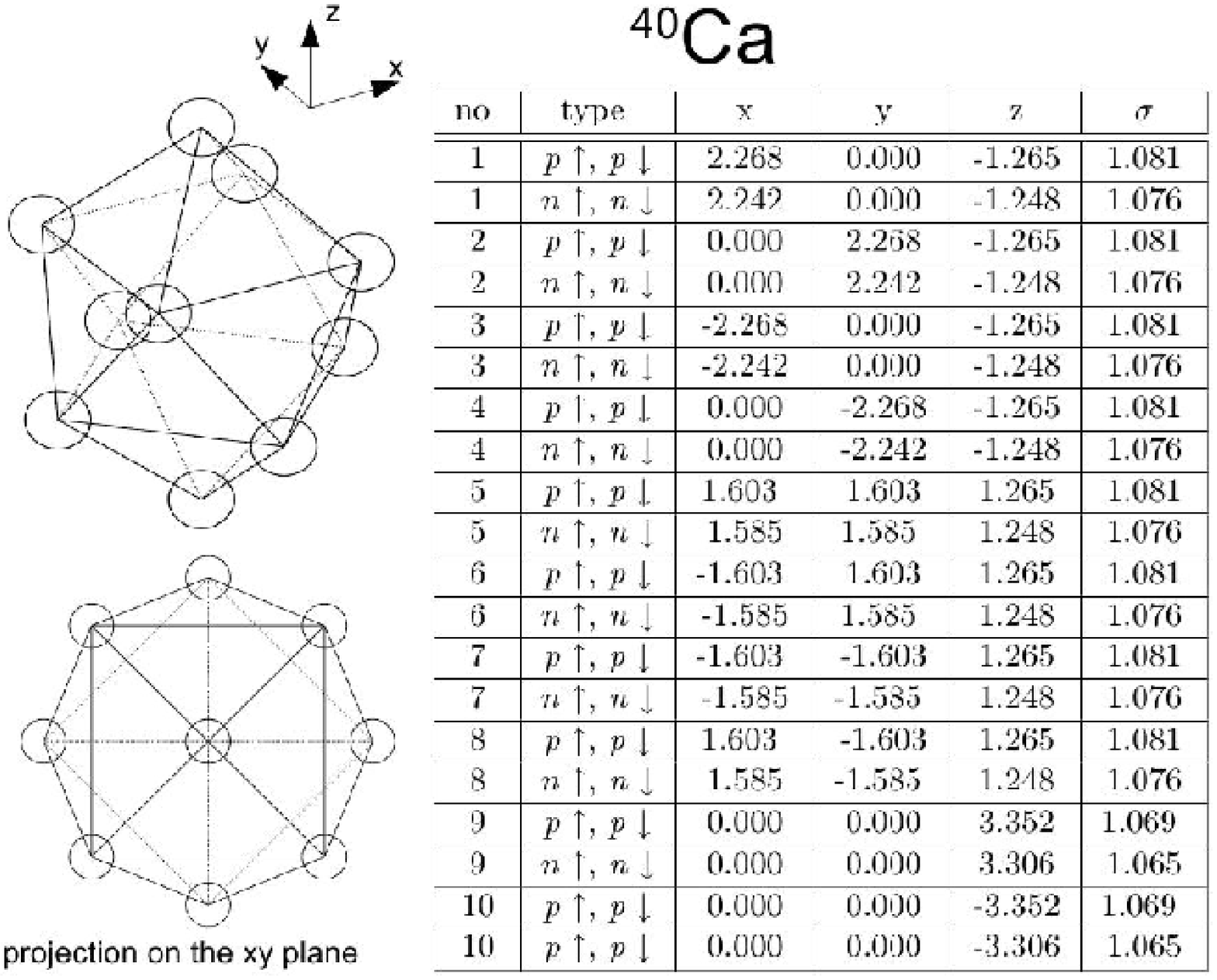}

\caption{The position and the variance of the wave packet for the nucleons
in the nucleus $^{40}Ca$. Notation is the same as in Figure 3.}
\end{figure}

The alpha structures obtained in our calculations are identical with
structures presented in papers \cite{Hafs1938}, \cite{Freer2010}
up to the silicon and the number of alpha-alpha bonds are the same
as that proposed by \cite{Bethe1936}(after $^{16}O$ there is an increment
of three bonds for each additional alpha particle). Thus we have 15
bonds for $^{28}Si$ instead of 16 \cite{Hafs1938}. The justification
\textbf{ } for this selection can be
understood from Fig. 7 for alpha-alpha distance distribution. 

\begin{figure}[h]
\includegraphics[width=12cm]{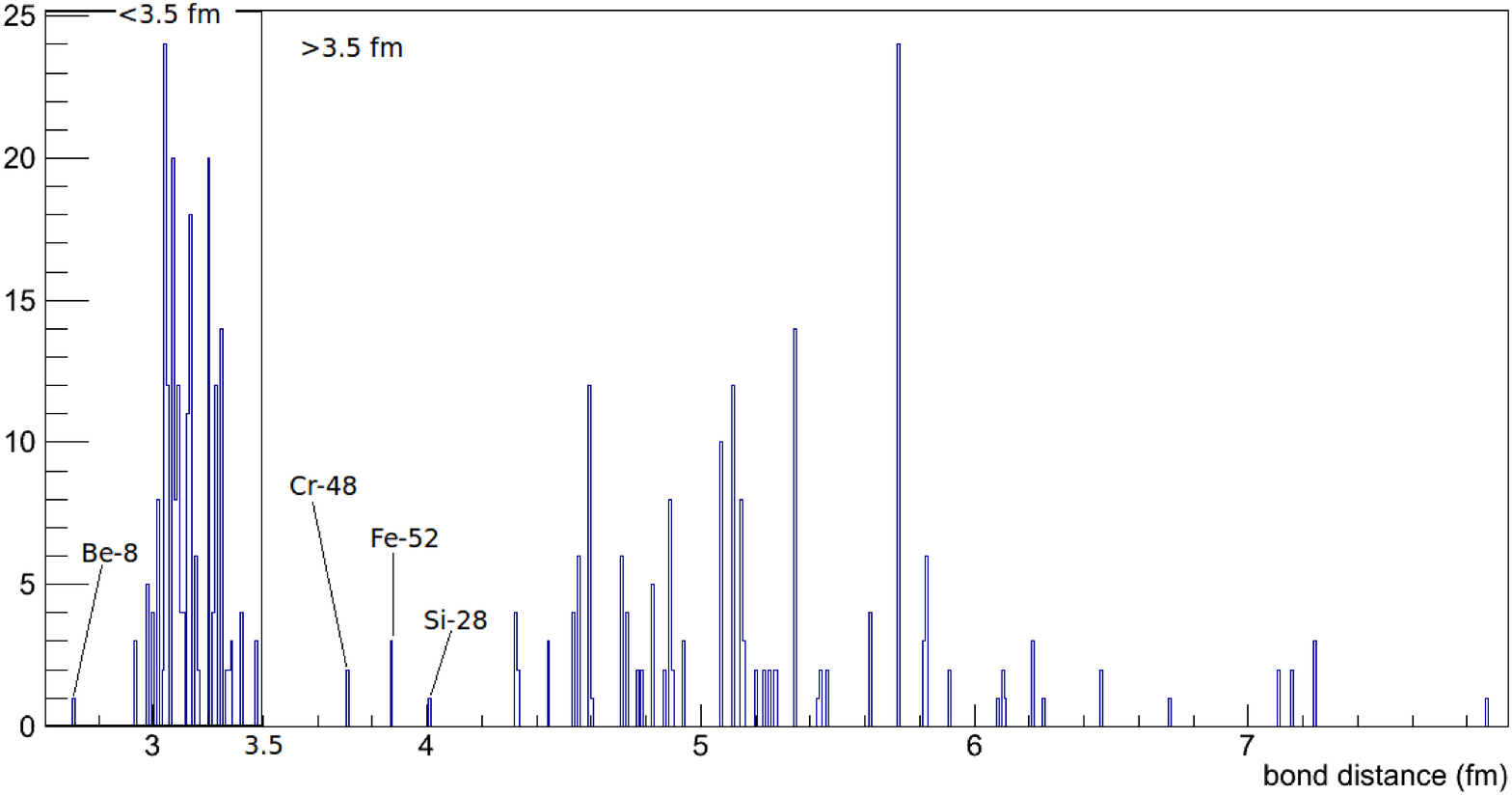}

\caption{The distribution alpha-alpha inter-particle distance calculated with
MLDM is plotted (up to $^{60}Zn$). Most of the countable distance
lies below 3.5 fm (not a final conclusion). For $^{28}Si$, $16^{th}$
bond length is shown.}
\end{figure}
This distribution presented in Fig. 6, shows that the countable alpha-alpha bond distances
are centred around $3.23 fm$ up to $3.5 fm$. In the case of $^{28}Si$,
the distances associated with 15 bonds lie below this value and the
rest has distances larger than $4.0 fm$. In our model calculation,
a systematic increase in the binding energy can be observed up to
$^{56}Ni$. For small $Z$ we obtained the structures identical to that
of Bethe's prediction \cite{Bethe1936}. The model results
also indicate that the first nucleus for which the change of systematic
increase in the number of bonds with an increase in the number of
alpha particles in the structure is $^{60}Zn$ (see Fig 8). This results
in a departure from the systematic increase in energy presented in
the Fig 9.

\begin{figure}[h]
\includegraphics[width=12cm]{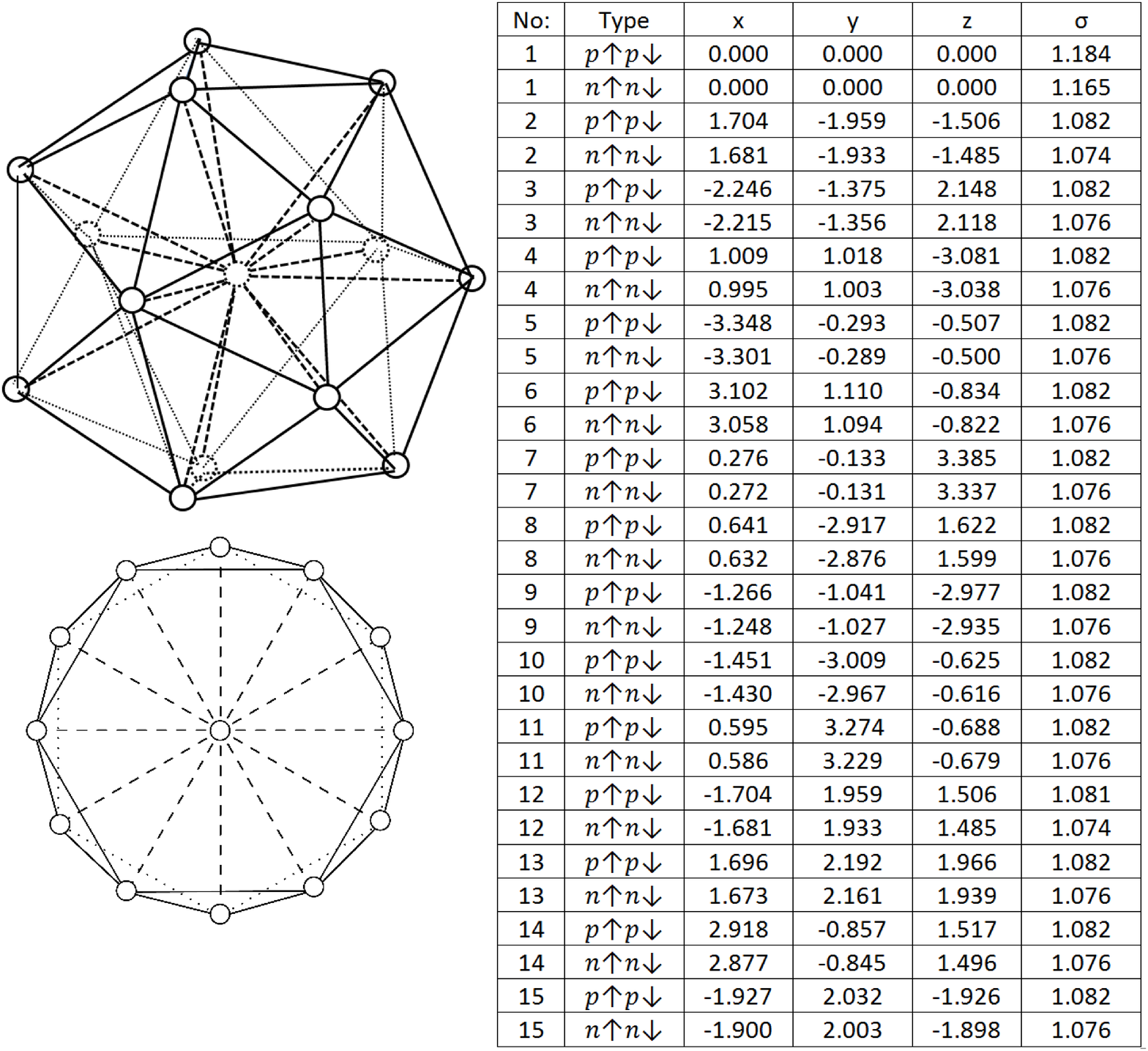}

\caption{The alpha particle structure of $^{60}Zn$ obtained via MLDM calculation
(the systematic shift in the increase of the binding energy starts
from this element, see the text for detail).}
\end{figure}

\begin{figure}[h]
\includegraphics[width=12cm]{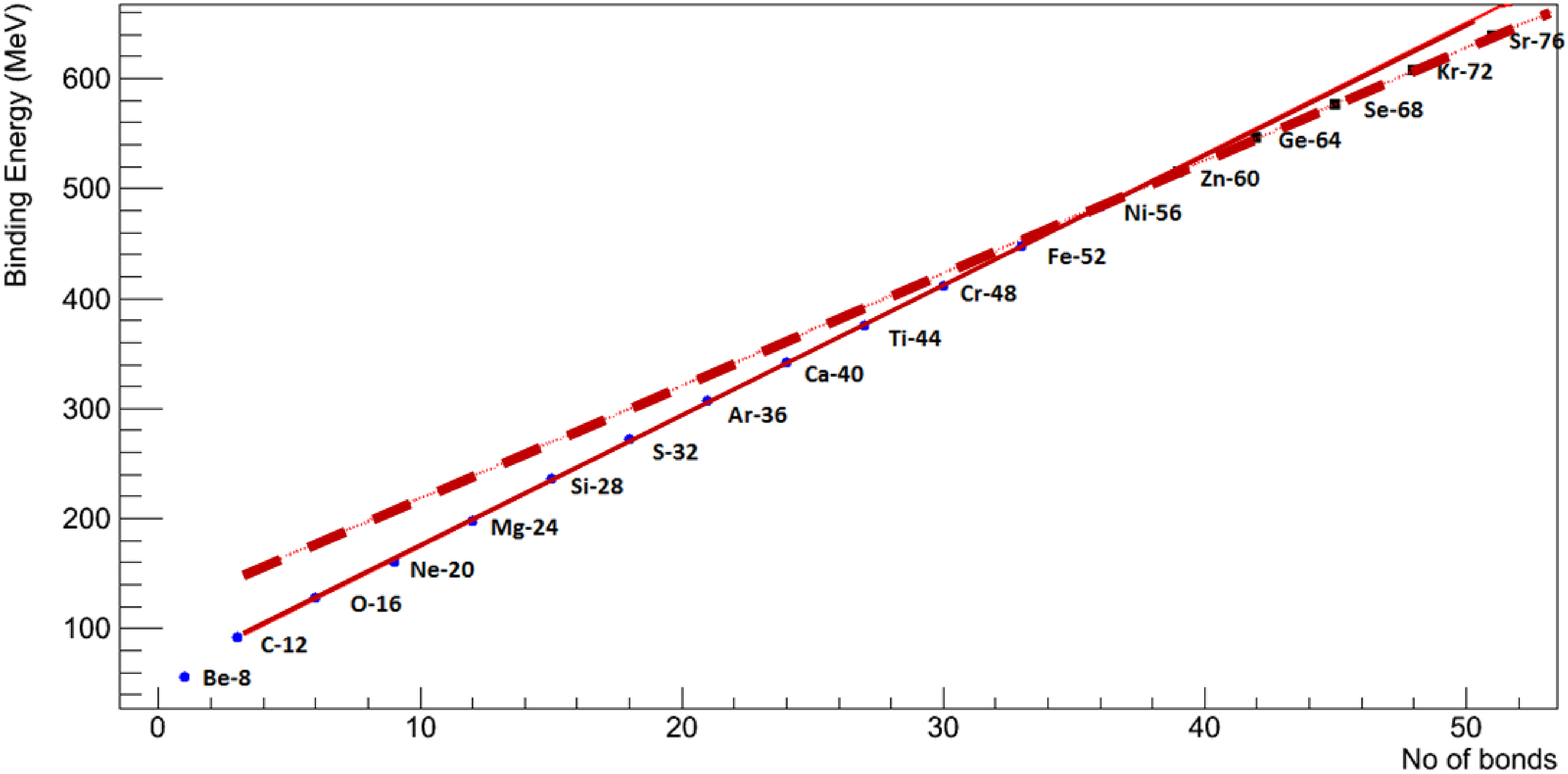}

\caption{Total binding energy of the nuclei versus number of bonds. From $^{12}C$
to $^{56}Ni$ , it follows the same linear relationship, according
to obtained structures. From $^{60}Zn$ , slope has a new linear relationship
reflecting change occurring in the nuclei structures, (see text for
details).}
\end{figure}

\section{Conclusions }

A new form of the EOS suitable for the MLDM model calculations is
presented. Preliminary results show that the MLDM model is able to
reproduce the basic properties of atomic nuclei. We see also that
MLDM describes fairly well the properties of very light nuclei. Because
light nuclei are usually manufactured with the highest probability
this is important for a correct description of the reaction dynamics.

Another important property of the model, resulting from taking into
account the mutual interaction related to the spin and isospin of
nucleons, is the emergence of the alpha structures which is particularly
noticeable for even-even nuclei with equal numbers of protons and
neutrons. This formation of alpha clusters is in line with considerations
of the binding energy gain associated with the resulting increase
in the number of bonds between alpha particles. It seems that the
model can be helpful in explaining of the change of the binding energy
increase occurring close to the nucleus $Z>28$. This creates an
additional branch in the scheme of the binding energy growth as can
be seen in the experimental systematics. 

A detailed analysis of the binding energy (as energy per nucleon)
and analysis of the nuclear density profiles show that the structures
described herein for certain nuclei may vary. This is due to the existence
of various structures that minimize the energy of the ground state.
For some nuclei the binding energy differences related to the various
structures are sometimes very small and subtle effects can lead to
selection of one of them. This also shows that the number of alpha
bonds may differ slightly in each individual case. The analysis of
this behaviour as well as restrictions on the applicability of the
present simple model will be presented in forthcoming paper.

We need to emphasize the very preliminary character of these results.
In order to predict better the values of the EOS parameters further
work is needed and we have to perform a global search in which we
will take into account all available experimental data. It would also
help to compare the model results, describing the basic observables
for heavy ion reactions (particularly for the compressibility factor
$K_{0}$) in the low energy region where the impact of the nucleon-nucleon
collisions can be neglected. 

\section*{Acknowledgement }

The authors are indebted to Professor L. Jarczyk and Professor R.
Płaneta for reading the manuscript and a fruitful discussion.

This work was supported in part by the Foundation for Polish Science
– MPD program, co-financed by the European Union within the European
Regional Development Fund. 

This work was supported also by the IN2P3 grant number 08-128.

\section*{Appendix A}

In this appendix the form of EOS was derived which is used in the
paper. We start from the expansion of the energy density function
and we use the symmetry which exists in nuclear interactions. According
to this symmetry, in the expansion (up to the $4th$ order) of the $e(\xi,\delta,\eta_{n},\eta{}_{p})$
around the point: $\xi=0,\:\delta=0,\:\eta_{n}=0,\:\eta_{p}=0$ only
the following terms appear:
\[
e=e_{0}+\frac{1}{2}e_{\xi\xi}\xi^{2}+
\]
\[
\delta^{2}\left(\frac{1}{2}e_{\delta\delta}+\frac{1}{4}e_{\delta\delta S_{n}S_{n}}\eta_{n}^{2}+\frac{1}{2}e_{\delta\delta S_{n}S_{p}}\eta_{n}\eta_{p}+\frac{1}{4}e_{\delta\delta S_{p}S_{p}}\eta_{p}^{2}+\frac{1}{2}e_{\xi\delta\delta}\xi+\frac{1}{4}e_{\xi\xi\delta\delta}\xi^{2}\right)+
\]

\[
\eta_{n}^{2}\left(\frac{1}{2}e_{\eta_{n}\eta_{n}}+\frac{1}{24}e_{\eta_{n}\eta_{n}\eta_{n}\eta_{n}}\eta_{n}^{2}+\frac{1}{4}e_{\eta_{n}\eta_{n}\eta_{p}\eta_{p}}\eta_{p}^{2}+\frac{1}{2}e_{\xi\eta_{n}\eta_{n}}\xi+\frac{1}{4}e_{\xi\xi\eta_{n}\eta_{n}}\xi^{2}\right)+
\]
\[
\eta_{p}^{2}\left(\frac{1}{2}e_{\eta_{p}\eta_{p}}+\frac{1}{24}e_{\eta_{p}\eta_{p}\eta_{p}\eta_{p}}\eta_{p}^{2}+\frac{1}{2}e_{\xi\eta_{p}\eta_{p}}\xi+\frac{1}{4}e_{\xi\xi\eta_{p}\eta_{p}}\xi^{2}\right)+
\]
\begin{equation}
\eta_{n}\eta_{p}\left(e_{\eta_{n}\eta_{p}}+e_{\xi\eta_{n}\eta_{p}}\xi+\frac{1}{2}e_{\xi\xi\eta_{n}\eta_{p}}\xi^{2}\right)\label{eq:eos_1}
\end{equation}

The symbol $e$ with the index constituting one of the variables of
the equation denotes a derivative of the energy density with respect
to this variable. We neglect the dependence on $\delta^{4}$ and for
small spin polarization the terms in brackets containing $\eta_{n}^{2},\:\eta_{p}^{2},\:\eta_{n}\eta_{p}$
can be neglected in relation to $e_{\delta\delta}$, $e_{\eta_{n}\eta_{n}}$,
$e_{\eta_{p}\eta_{p}}$ , and finally one can write: 
\[
e=e_{0}+\frac{1}{2}e_{\xi\xi}\xi^{2}+
\]
\[
\delta^{2}\left(\frac{1}{2}e_{\delta\delta}+\frac{1}{2}e_{\xi\delta\delta}\xi+\frac{1}{4}e_{\xi\xi\delta\delta}\xi^{2}\right)+
\]

\[
\eta_{n}^{2}\left(\frac{1}{2}e_{\eta_{n}\eta_{n}}+\frac{1}{2}e_{\xi\eta_{n}\eta_{n}}\xi+\frac{1}{4}e_{\xi\xi\eta_{n}\eta_{n}}\xi^{2}\right)+
\]
\[
\eta_{p}^{2}\left(\frac{1}{2}e_{\eta_{p}\eta_{p}}+\frac{1}{2}e_{\xi\eta_{p}\eta_{p}}\xi+\frac{1}{4}e_{\xi\xi\eta_{p}\eta_{p}}\xi^{2}\right)+
\]
\begin{equation}
2\eta_{n}\eta_{p}\left(\frac{1}{2}e_{\eta_{n}\eta_{p}}+\frac{1}{2}e_{\xi\eta_{n}\eta_{p}}\xi+\frac{1}{4}e_{\xi\xi\eta_{n}\eta_{p}}\xi^{2}\right)\label{eq:eos_2}
\end{equation}
Replacing symbols of derivatives $e_{\lambda...}$by variables defined
in the following relations: 
\[
\frac{K_{0}}{18}=\frac{1}{2}e_{\xi\xi}
\]
\[
e_{I0}=\frac{1}{2}e_{\delta\delta}\,,\;\frac{L_{I}}{3}=\frac{1}{2}e_{\xi\delta\delta}\,,\;\frac{K_{I}}{18}=\frac{1}{4}e_{\xi\xi\delta\delta}
\]
\[
e_{ii0}=\frac{1}{2}e_{\eta_{n}\eta_{n}}\,,\;\frac{L_{ii}}{3}=\frac{1}{2}e_{\xi\eta_{n}\eta_{n}}\,,\;\frac{K_{ii}}{18}=\frac{1}{4}e_{\xi\xi\eta_{n}\eta_{n}}
\]
\begin{equation}
e_{ij0}=\frac{1}{2}e_{\eta_{n}\eta_{p}}\,,\;\frac{L_{ij}}{3}=\frac{1}{2}e_{\xi\eta_{n}\eta_{p}}\,,\;\frac{K_{ij}}{18}=\frac{1}{4}e_{\xi\xi\eta_{n}\eta_{p}}\label{eq:eos_3}
\end{equation}
and taking into account the mentioned symmetries of nuclear interactions
(the derivatives $\left(e_{\eta_{p}\eta_{p}},\: e_{\eta_{n}\eta_{n}}\right)$,
$\left(e_{\xi\eta_{p}\eta_{p}},\: e_{\xi\eta_{n}\eta_{n}}\right)$
and $\left(e_{\xi\xi\eta_{p}\eta_{p}},\: e_{\xi\xi\eta_{n}\eta_{n}}\right)$
are pairwise equal,and this means that the respective symmetry energy
associated with the spin is identical for protons and neutrons), we
obtain the equation (\ref{eq:eos_4}).

\section*{Appendix B}

In this appendix we present the method of calculation of the variance
of energy given by the EOS in the neighbourhood of the center of the
packet $k$. In the first step we calculate the average energy density
of matter at random points with the distribution of $^{k}\!\rho(\mathbf{r})$.
\begin{equation}
\!\bar{e}_{k}(k)=\intop e(\rho,\delta,\eta_{n},\eta_{p})\cdot^{k}\!\rho(\mathbf{r})d^{3}\mathbf{r}\label{eq:ave_e_k}
\end{equation}
If we use the equation of state in the form of an expansion (\ref{eq:eos_4})
and (\ref{eq:form}) for the isospin part the derived term cannot
be solved analytically:

\begin{equation}
\intop\delta^{2}\alpha_{I}\rho\cdot^{k}\!\!\!\rho(\mathbf{r})d^{3}\mathbf{r=\alpha_{I}\intop\left(\frac{\rho_{\mathit{n}}-\rho_{\mathit{p}}}{\rho}\right)^{2}\rho\cdot^{k}\!\rho\mathit{(\mathbf{r})d^{3}}r=\alpha_{I}\intop\mathit{\frac{\rho_{\mathit{n}}\rho_{\mathit{n}}-\mathit{2\rho_{n}\rho_{p}}+\rho_{p}\rho_{p}}{\rho}\cdot^{k}}\!\rho\mathit{(\mathbf{r})d}^{3}r}\label{eq:d2_alf_I}
\end{equation}
In order to avoid this problem we can use the following approximations:
\begin{equation}
\alpha_{I}\intop\frac{\rho_{\mathit{n}}\rho_{n}}{\rho}\cdot^{k}\!\rho(\mathbf{r})d^{3}\mathbf{r}\cong\alpha_{I}\frac{N}{A}\intop\rho_{\mathit{n}}\cdot^{k}\!\rho(\mathbf{r})d^{3}\mathbf{r}\label{eq:app_roro_ro}
\end{equation}
where $N$ and $A$ are the total number of neutrons and nuclei in
the system, respectively. As we can see, after removing $\rho$ from
the denominator the integral can be solved analytically. Similarly,
we can find the remaining ingredients (\ref{eq:d2_alf_I}) (with $\rho_{n}\rho_{p}$
and $\rho_{p}\rho_{p}$ ). 

In the next step we assume that the $\delta,\:\eta_{n},\:\eta_{p}$
are established in the vicinity of the nucleon $k$ as $\bar{\delta}(k),\:\bar{\eta}_{n}(k),\:\bar{\eta}_{p}(k)$
(we replace them by the corresponding average values) and we determine
the effective density of $\rho_{ef}(k)$ based on the equation:
\begin{equation}
\!\bar{e}_{k}=e\left(\rho_{ef}(k),\bar{\delta}(k),\bar{\eta}_{n}(k),\bar{\eta}_{p}(k)\right)\label{eq:ave_eq_e}
\end{equation}
In further considerations we replace the set $\rho_{ef}(k),\:\bar{\delta}(k),\:\bar{\eta}_{n}(k),\:\bar{\eta}_{p}(k)$
by $\tilde{\rho}_{p\uparrow}(k),\:\tilde{\rho}_{p\downarrow},\:\tilde{\rho}_{n\uparrow}(k),\:\tilde{\rho}_{n\downarrow}(k)$.
To calculate the variance $\sigma_{k}^{2}(e)$ we develop the integral
expression: 
\begin{equation}
\sigma_{k}^{2}(e)=\intop\left(\bar{e}_{k}-e(\rho_{p\uparrow},\rho_{p\downarrow},\rho_{n\uparrow},\rho_{n\downarrow})\right)^{2}\cdot^{k}\!\!\!\rho(\mathbf{r})d^{3}\mathbf{r}\label{eq:sig2e_k}
\end{equation}
around $\tilde{\rho}_{p\uparrow}(k),\:\tilde{\rho}_{p\downarrow},\:\tilde{\rho}_{n\uparrow}(k),\:\tilde{\rho}_{n\downarrow}(k)$.
This allows the analytical calculation of the the considered integral.
In our calculations we limited ourselves to the first of order expansion.
So, because  $e\left(\tilde{\rho}_{p\uparrow}(k),\tilde{\rho}_{p\downarrow}(k),\tilde{\rho}_{n\uparrow}(k),\tilde{\rho}_{n\downarrow}(k)\right)=\bar{e}_{k}$
this choice of expansion gives a very simple form of the integrand.

\end{document}